\documentclass[10pt]{iopart}
\usepackage{amssymb}

\begin{document}

\title[Brans-Dicke gravity and capture]{Brans-Dicke gravity and the capture of stars by black holes:
some asymptotic results}
\author{Bertrand Chauvineau$^{1}$\footnote {Member of CNAP, Comit\'e National des Astronomes et Physiciens}, Alessandro D.A.M. Spallicci$^{2}$\footnote{Associated member of CNRS, Centre National de la Recherche Scientifique} and Jean-Daniel
Fournier$^{2}$\footnote{Member of CNRS, Centre National de la Recherche Scientifique }}

\address{{ UMR 6162 D\'ept. d'Astrophysique Relativiste ARTEMIS}\\
{CNRS, Observatoire de la C\^{o}te d'Azur, France}\\
{$^1$ Avenue Copernic, 06130 Grasse }\\
{$^2$ Bd de l'Observatoire, 06304 Nice }}

\begin{abstract}
In the context of star capture by a black hole, a new noticeable difference between Brans-Dicke theory and general relativity gravitational radiation is pointed out. This feature stems from the 
non-stationarity of the black hole state, barring Hawking's theorem.
\end{abstract}

\bigskip

Emails : chauvineau@obs-azur.fr, spallicci@obs-nice.fr, fournier@obs-nice.fr

\bigskip

Pacs numbers : 04.50.+h, 04.30.Db


\section{Introduction}

General Relativity (GR) predicts radiation that, in the lowest order, is
proportional to the third derivative of the quadrupole moment of the
mass-energy distribution. It is a consequence of the conservation equations
that the first derivative of the monopole moment and the second derivative
of the dipole moment are zero. In Brans-Dicke (BD) theory (Brans and Dicke, 1961; Weinberg, 1972; Will, 1993) the scalar field $\Phi$ determines the local value of 
the gravitation ''constant'' G. A known feature of BD theory consists of the scalar field variability
that generates contributions of dipolar terms to the gravitational radiation.

Herein, we show a new noticeable feature of BD theory in a non-stationary regime, which could happen after the capture
of a star by a black hole.
The BD vacuum field equations are 
\begin{equation}
R_{\alpha \beta }=\frac{1}{\Phi }\nabla _{\alpha }\partial _{\beta }\Phi +%
\frac{\omega }{\Phi ^{2}}\partial _{\alpha }\Phi \partial _{\beta }\Phi ,
\label{eq:1}
\end{equation}
\begin{equation}
g^{\alpha\beta}\nabla_{\alpha}\partial_{\beta}\Phi =0 ,  \label{eq:2}
\end{equation}
where $\Phi$ is the scalar field and $\omega$ its coupling to matter.

Since we are concerned in the post-capture phase, we write eq.(\ref{eq:2}) without source term. This does not imply 
that eq.(\ref{eq:2}) admits only trivial solutions; for instance, it allows solutions corresponding to a non vanishing 
gradient of the scalar field. 
Hawking's theorem (1972) on BD black holes, later generalised to the case of a
scalar tensor black hole (Bekenstein and Meisels, 1978) states that under the following
hypotheses:

(h1) : stationarity,

(h2) : asymptotical flatness,

(h3) : vacuum,

(h4) : existence of an event horizon $\left( H\right) $,

(h5) : finiteness of the area of $\left( H\right) $,

(h6) : bounded value of $\sqrt{\left| 2\omega +3\right| }~\varphi ~\partial
\varphi $ on $\left( H\right) $, where $\varphi =\Phi -\Phi _{0}$ ($\Phi
_{0} $ being the asymptotic value of $\Phi $ at spatial infinity),\newline
the solution of the BD field equations is a GR solution, i.e $\varphi \equiv 0$ everywhere; 
thus the common saying on BD black holes being identical to
GR ones. We stress the importance of the concurrence of all the
previous hypothesis, (h1-h6). Indeed, there are BD stationary black holes
solutions which are not GR solutions due to the breakdown either of the 
hypothesis (h5) (Bronnikov et al., 1998) or (h2) (Singh and Rai, 1979;
Kim, 1999).

Relaxing the hypothesis (h1), non-stationary BD black holes generally differ
from the GR ones. In the present paper, we are interested in the problem
of black hole relaxation after the capture of a star, in the framework of BD
theory. As usual, we assume that the black hole under consideration satisfies
the hypotheses (h2), (h5) and (h6). After the star has been captured, and
during the relaxation phase, the hypothesis (h3) is assumed to be
satisfied too. (h4) is naturally satisfied, since the endproduct is a black
hole. Then, the above theorem ensures that the final (stationary) state
is necessarily a GR black hole solution.

\section{The capture scenario in Brans-Dicke theory}

After the capture of a star, the black hole undergoes oscillations and
relaxes towards a GR stationary state. Thus, adopting a perturbative approach,
we rewrite the metric and the scalar field as

\begin{equation}
g_{\alpha \beta }=\sigma_{\alpha \beta }+h_{\alpha \beta }
\label{3}
\end{equation}
\begin{equation}
\Phi =1+\varphi ,  \label{4}
\end{equation}
where $\sigma_{\alpha \beta }$ is a GR black hole metric
(Schwarzschild or Kerr), and $\Phi _{0}=1$ is the normalized value of the
scalar field at infinity, after relaxation. $\left| h_{\alpha \beta }\right| 
$, $\left| \partial h_{\alpha \beta }\right| $, $\left| \varphi \right| $
and $\left| \partial \varphi \right| $\ are assumed to be small enough to
ensure the applicability of the perturbative approach. Besides, 
$h_{\alpha \beta }$, $\varphi $ and their derivatives vanish in the large time limit, since the
final state is the GR solution.

We assume that $\partial \varphi $ and $r_{BH}\partial \partial \varphi $
have the same order of magnitude, where $r_{BH}$ is a typical ''length''
associated to the black hole (for instance, the Schwarzschild radius, in the
spherical case).

To lowest order, eq. (\ref{eq:2}) reads 
\begin{equation}
\partial _{\alpha }\left( \sqrt{-\sigma}~\sigma^{\alpha \beta }\partial _{\beta }\varphi \right) =0.  \label{6}
\end{equation}
Then, $\varphi $ decouples from $h_{\alpha \beta }$, and the integration of (%
\ref{6}) gives, in principle, $\varphi $ from $\sigma
_{\alpha \beta }$. In eq. (\ref{eq:1}), the Ricci tensor reads $%
R_{\alpha \beta }\left( g_{\mu \nu }\right) =R_{\alpha \beta }\left(
\sigma_{\mu \nu }\right) +\pounds _{\alpha \beta }\left(
h_{\mu \nu }\right) =\pounds _{\alpha \beta }\left( h_{\mu \nu }\right) $,
since $\sigma_{\mu \nu }$ satisfies the GR vacuum equation 
$R_{\alpha \beta }=0$. The linear operator $\pounds $ depends on the
GR final state only, i.e. on its parameters (mass, angular momentum).

An order of magnitude estimate of the r.h.s. terms of eq. (\ref{eq:1}) shows that $\Phi ^{-1}\nabla _{\alpha
}\partial _{\beta }\Phi \sim {\bar \nabla} _{\alpha
}\partial _{\beta }\Phi \sim r_{BH}^{-1}\partial \varphi $, and $\omega \Phi
^{-2}\partial _{\alpha }\Phi \partial _{\beta }\Phi \sim \omega \partial
_{\alpha }\varphi \partial _{\beta }\varphi \sim \omega \left( \partial
\varphi \right) ^{2}$, where the operator ${\bar\nabla} $
is the covariant derivative with respect to the background metric $%
\sigma$. The term $\omega \left( \partial \varphi \right)
^{2}$ is a second order term, but it is weighted by the BD parameter $\omega 
$, known to be very large. Then, even when $\left| \partial \varphi
\right| $ is $\ll 1$ (but finite), the quadratic term can be comparable or
greater than the linear one. Finally, eq. (\ref{eq:1}) reads 
\begin{equation}
\pounds _{\alpha \beta }\left( h_{\mu \nu }\right) =S_{\alpha \beta },
\label{10}
\end{equation}
with 
\begin{eqnarray}
S_{\alpha \beta } &\simeq &\omega \partial _{\alpha }\varphi \partial
_{\beta }\varphi ~~~\mathrm{when}~~~\left| \partial \varphi \right| >\frac{1%
}{\omega r_{BH}}~~~\mathrm{(quadratic~phase),}  \label{11} \\
&\simeq &{\nabla }_{\alpha }\partial _{\beta }\varphi ~~~\mathrm{when}%
~~~\left| \partial \varphi \right| <\frac{1}{\omega r_{BH}}~~~\mathrm{%
(linear~phase).}
\end{eqnarray}
\qquad The quadratic phase occurs, or better {\it may occur}, just after the star
absorption by the black hole. Conversely, the linear phase takes \textit{always%
} place, before stationarity (unless the whole solution is a GR one, i.e. is
such that $\Phi =$ constant). The quadratic phase doesn't necessarily occur. 
Indeed, solutions of eqs. (\ref{6},\ref{10}) exist for which $\partial \varphi $ is
always smaller than $1/\left( \omega r_{BH}\right) $: 
an obvious example is provided by GR black hole oscillations,
corresponding to the case $\varphi = 0$ everywhere, a particular case of BD
black hole oscillations.\\
But, if the initial values of $\partial
\varphi $ are sufficiently large, such that the quadratic phase does take
place, then the transition towards the linear phase \textit{does necessarily
exist}, due to the black hole acquisition of a GR stationary state at large times. 
At this stage, we can thus raise two questions :

(q1) : does such a transition, from the quadratic to the linear regime, in $%
\partial \varphi $, of the source, have a {\it characteristic} signature on $%
h_{\alpha \beta }$ ?

(q2)\ :\ is this signature essentially {\it non dependent} upon the precise form
of the linear operator $\pounds _{\alpha \beta }$ ?

\section{Discussion on the Schwarzschild case}

We consider the case where the final state is a Schwarzschild black hole. In
this case, the scalar field equation (\ref{6}) reads 

\begin{equation}
\fl
\frac{-r^{2}}{1-r_{g}/r}\frac{\partial ^{2}\varphi }{\partial t^{2}}+\frac{%
\partial }{\partial r}\left [\left( r^{2}-r_{g}r\right) \frac{\partial
\varphi }{\partial r}\right ] +\frac{1}{\sin \theta }\frac{\partial }{%
\partial \theta }\left( \sin \theta \frac{\partial \varphi }{\partial \theta 
}\right) +\frac{1}{\sin ^{2}\theta }\frac{\partial ^{2}\varphi }{\partial
\phi ^{2}}=0,  \label{14}
\end{equation}
where $r_{g}$ is the Schwarzschild radius.

The variables separation technique leads to the following particular solutions 
\begin{equation}
\varphi \equiv e^{-\lambda t}R(r)Y(\theta ,\phi ),  \label{15}
\end{equation}
where $\lambda $ is a positive separation constant and $Y(\theta ,\phi )$ any
spherical harmonic function. The radial function $R(r)$ has the following asymptotic forms 
\begin{equation}
R(r\simeq r_g) \simeq c_{1}\left( \frac{r}{r_{g}}-1\right) ^{\lambda r_{g}}
\label{16}
\end{equation}

\begin{equation}
R(r \gg 1)\simeq c_2 \frac{e^{-\lambda r}}{r}
\label{17}
\end{equation}
where $c_{1,2}$\ are constants. It is noteworthy that the exponent in eq. (\ref{16}) and the 
decrement in eq. (\ref{17}) are expressed in terms of the constant $\lambda$, previously introduced as time decrement 
in eq. (\ref{15}).

As consequence of the latter, the time-dependence of the source term in (\ref{10}) is given by 
\begin{eqnarray}
S_{\alpha \beta } &\sim &e^{-2\lambda t}~~~\mathrm{(quadratic~phase)}
\label{18} \\
&\sim &e^{-\lambda t}~~~\mathrm{(linear~phase).} 
\end{eqnarray}
The general solution of (\ref{10}) reads then 
\begin{equation}
h_{\alpha \beta }=h_{\alpha \beta }^{\left( H\right) }+h_{\alpha \beta
}^{\left( P\right) },  \label{19}
\end{equation}
where $h_{\alpha \beta }^{\left( H\right) }$\ is the general solution of $%
\pounds _{\alpha \beta }\left( h_{\mu \nu }\right) =0$ (GR case), and $%
h_{\alpha \beta }^{\left( P\right) }$\ is a particular solution of the
complete eq. (\ref{10}). Since the linear operator $\pounds _{\alpha
\beta }$ is time-independent, $h_{\alpha \beta }^{\left( P\right) }$\ can be
constructed with the same time-dependence as $S_{\alpha \beta }$.

In those cases, where $h^{\left( P\right) }$ is not very small with
respect to $h^{\left( H\right) }$, this will result into a change in the
decrement of the signal, as direct consequence of the transition from the $%
e^{-2\lambda t}$ to the $e^{-\lambda t}$\ dependence in the source term.
This decrement change, solely due to the presence of the BD scalar field, is to
be perceived as a qualitative difference between BD and GR black hole
oscillations.

\section{Conclusions}

We have identified a transition between two successive regimes during the radiation 
emission after capture of a star by a black hole. While such transition is obviously absent in 
general relativity, conversely it may be present in Brans-Dicke and concurrent scalar-tensor theories of gravity (Will, 1993).

\section{Acknowledgments}

The European Space Agency is acknowledged for granting the G. Colombo Senior
Fellowship to A. Spallicci.

\bigskip

\section*{References}

Bekenstein J D and Meisels A 1978 Phys. Rev. D \textbf{18} 4378 \newline
Brans C and Dicke R H 1961 Phys Rev 124 925 \newline
Bronnikov K A Clement G Constantinidis C P and Fabris J C 1998 Phys. Lett. A 
\textbf{243} 121 \newline
Hawking S W 1972 Commun. Math. Phys. \textbf{25} 167 \newline
Kim H 1999 Phys. Rev. D \textbf{60} 024001 \newline
Singh T Rai L N 1979 Gen. Rel. Grav. \textbf{11} 37\newline
Weinberg S 1972 {\it Gravitation and cosmology: principles and applications of the general theory of relativity}
John Wiley \& Sons\newline 
Will C M 1993 {\it Theory and experiments in gravitational physics} Cambridge Univ. Press
\end{document}